\documentclass[prl]{revtex4}

\usepackage{graphicx}
\usepackage{dcolumn}
\usepackage{amsmath}
\usepackage{epsfig}
\usepackage{amssymb}

\makeatletter
\def\btt#1{\texttt{\@backslashchar#1}}
\DeclareRobustCommand\bblash{\btt{\@backslashchar}}
\makeatother

\def\Bid{{\mathchoice {\rm {1\mskip-4.5mu l}} {\rm
{1\mskip-4.5mu l}} {\rm {1\mskip-3.8mu l}} {\rm {1\mskip-4.3mu l}}}}

\begin{document}


\title{Combined Error Correction Techniques for Quantum Computing 
Architectures}

\author{Mark S. Byrd}
\email{mbyrd@chem.utoronto.ca}
\author{Daniel A. Lidar}
\email{dlidar@chem.utoronto.ca}

\affiliation{Chemical Physics Theory Group, University of Toronto, 
80 St. George Street, Toronto, Ontario M5S 3H6, Canada}


\begin{abstract}
Proposals for quantum computing devices are many and varied.  They 
each have unique noise processes that make none of them fully reliable 
at this time.  There are several error correction/avoidance 
techniques which are valuable for reducing or eliminating errors, 
but not one, alone, will serve as a panacea.  One must 
therefore take advantage of the strength of each of these 
techniques so that we may extend the coherence times of the 
quantum systems and create more reliable computing devices.  
To this end we give a general strategy for using
dynamical decoupling operations on encoded
subspaces. These encodings may be of any form; 
of particular importance are decoherence-free 
subspaces and quantum error correction codes.  We then give 
means for empirically determining an appropriate set of 
dynamical decoupling operations for a given experiment.  
Using these techniques, we then
propose a comprehensive encoding solution to many of the problems of 
quantum computing proposals which use exchange-type interactions.  
This uses a decoherence-free subspace and an efficient set of 
dynamical decoupling operations.  It also addresses the problems
of controllability in  
solid state quantum dot devices.  
\end{abstract}


\maketitle



\section{Quantum Error Correction Strategies}

The main obstacle to building a quantum computing device is 
noise and decoherence in the quantum system 
due to the inevitable interaction with the environment.  
There are several error 
correction/avoidance strategies for treating this problem.  They 
can be divided into three broad categories
Quantum error correction codes (QECCs) 
\cite{Shor:95,Steane:96a,Gottesman:97,Knill:97b}, 
(for a review see \cite{Steane:99}) use redundancy and an active 
measurement and recovery scheme to correct errors that occur 
during a computation (we include
topological quantum codes in this category; see \cite{Preskill:99} and
references therein).  Decoherence-free subspaces (DFSs) and noiseless
subsystems 
\cite{Zanardi:97c,Duan:98,Lidar:PRL98,Knill:99a,Kempe:00,Lidar:00a}, rely on symmetric 
system-bath interactions to find encodings that are immune 
to decoherence effects.  Dynamical decoupling, or ``bang-bang'' (BB) 
operations
\cite{Viola:98,Duan:98e,Viola:99,Zanardi:98b,Vitali:99,Viola:99a,Viola:00a,Vitali:01,ByrdLidar:01,Cory:00,Agarwal:01,Uchiyama:02}
are strong and fast pulses which suppress 
errors by averaging them away.  QECCs use extra qubits, which, 
at this time are a scarce resource.  They require at least a 
5 physical qubit to 1 logical qubit encoding 
\cite{Knill:97b,Laflamme:96} (neglecting 
ancillas required for fault-tolerant recovery) in order to correct a
general single qubit error \cite{Gottesman:97}.  
DFSs also require extra qubits and are most effective for 
collective errors, or errors where multiple qubits are coupled to the
same bath mode \cite{Lidar:00a}.  
The minimal encoding for a single qubit undergoing collective
decoherence is 3 physical qubits to 
one logical qubit \cite{Knill:99a}.  Finally, the BB control 
method requires 
a complete set of pulses to be implemented within the correlation 
time of the bath \cite{Viola:98}.  It does not, however, necessarily 
require extra qubits.

In this article we discuss the combination of BB operations with 
other encoding techniques to conserve qubit resources 
while making quantum computing devices more robust. An experiment
combining BB with quantum error correction has recently been reported \cite{Boulant:02}. We begin 
with a brief review of the BB control formalism before 
presenting an empirical formula for the 
determination of BB operations from a set of quantum 
process tomography measurements \cite{Nielsen:book}.  
We then give a general theorem which provides sufficient conditions 
for the elimination of errors via BB controls on 
logically encoded subspaces.  This is used to discuss the 
application of the BB controls in conjuction with QECCs.  
Our results are then used to determine a combined effective encoding, recoupling 
\cite{LidarWu:01}, and decoupling (BB) strategy for quantum computing 
devices which rely on exchange-type interactions \cite{ByrdLidar:01a}.  
We give estimates for the number of BB operations 
that can be performed in experiments using spin-coupled quantum 
dots in GaAs.  The estimates are based upon models 
of the underlying mechanisms of decoherence in these systems.  
However, the empirical method for determining 
BB operations, proposed in \cite{ByrdLidar:01a,ByrdLidar:02}, 
circumvents the need for a detailed understanding 
of the underlying decoherence processes.


\subsection{Bang Bang Operations}

Let us briefly review some important aspects of the method of 
BB controls.  BB controls are strong and fast
pulses, applied cyclically, which average out the environment-induced noise 
\cite{Viola:98}. In the limit of infinitely fast pulsing, BB controls have
been shown to completely remove decoherence. The simplest example of BB is
the ``parity-kick'' sequence \cite{Viola:98,Vitali:99}. Suppose 
that an error $E$ (an operator in the system-bath Hamiltonian) 
acts on the system, and that we can 
find a pulse $U$ (unitary operator) that anticommutes with $E$, 
and therefore changes the sign of this error: 
\begin{equation}
\{E,U\}=0,\;\;\Rightarrow \;\;U^{\dagger }EU=-E.  \label{eq:PK}
\end{equation}
Allowing the system to repeatedly undergo the sequence: $\{$free evolution
under $E$ (for time $\Delta t$), application of $U$, free evolution, 
application of $U^{-1}\}$, will cause the error
to be averaged out (``symmetrized'' \cite{Zanardi:98b,Viola:99}), 
thus \emph{decoupling} system and bath. The parity kick (whose origins
can be traced to the well-known Carr-Purcell sequence of NMR \cite{Carr:54})
and its generalizations have been
the subject of several recent publications
\cite{Viola:98,Duan:98e,Viola:99,Zanardi:98b,Vitali:99,Viola:99a,Viola:00a,Vitali:01,ByrdLidar:01,Cory:00,Agarwal:01,Uchiyama:02}.
In reality, for decoupling to work the
time taken for a complete cycle of pulses, $T_{c}$, must be significantly
shorter than the fastest bath correlation time $\tau _{c}$: 
\begin{equation}
\Delta t\leq T_{c}\ll \tau _{c}.  \label{eq:times}
\end{equation}
Even in the case that the time scales are close, one can achieve some noise
reduction \cite{Viola:98,Duan:98a,Vitali:99,Vitali:01,Uchiyama:02}. 
Knowledge of $\tau_{c}$, the inverse of the bath spectral density high
frequency
cut-off, is clearly desirable for determining 
the success of the BB procedure, and will
be discussed in detail below for quantum dots. Given that pulses have finite
durations, the ratio $\tau _{c}/T_{c}$ imposes further constraints on the
length of the experimentally implementable pulse sequences. However the 
empirical method for the determination of the BB operations, 
outlined below, takes these constraints into account.  


\subsection{Empirically Determined BB Controls}

Previous analyses of BB controls have typically assumed 
model system-bath Hamiltonians 
\cite{Viola:98,Duan:98e,Viola:99,Zanardi:98b,Vitali:99,Viola:99a,Viola:00a,Vitali:01,ByrdLidar:01,Cory:00,Agarwal:01,Uchiyama:02}.
However, the total system-bath 
Hamiltonian is often not known.  
As an alternative to this model-based approach we review here a
procedure we have previously proposed
for finding BB operations from experimental data \cite{ByrdLidar:01a,ByrdLidar:02}.  
This empirical determination requires neither 
a detailed understanding of the fundamental processes nor a detailed 
experimental analysis of each of the decoherence processes in the system. It 
requires only a set of quantum process tomography measurements 
\cite{Nielsen:book} on the logical qubits to determine the {\emph{types}} of 
errors that occur.  With this, one may empirically determine 
the set of {\em required} corrective pulses and the efficacy of the 
{\em experimentally available} pulse set \cite{ByrdLidar:01}.

Empirical BB is based on the following set of observations.  
Very generally, the evolution of an open quantum system, 
described by a density matrix 
$\rho$, satisfies the (completely positive \cite{Kraus:83}) map 
\begin{equation}
\rho (t)=\sum_{\alpha ,\beta }\ \chi _{\alpha \beta }(t)K_{\alpha }\rho
(0)K_{\beta }^{\dagger },  \label{eq:OSR}
\end{equation}
where the matrix $\chi_{\alpha \beta }(t)$ is hermitian and 
$\{K_{\alpha }\}$ is a system operator basis 
\cite{Chuang:97c,Lidar:CP01}.  The $\chi$
matrix can be determined from a quantum process tomography measurement 
\cite{Chuang:97c}.  It can be shown that Eq.~(\ref{eq:OSR}) 
can be transformed into \cite{Bacon:99} (using $\hbar = 1$)
\begin{equation}
\label{eq:OSRexp}
\rho(t) = -i [S(t),\rho(0)] + \frac{1}{2}\sum_{\alpha,\beta = 1}
          \chi_{\alpha \beta}(t)\left([K_\alpha,\rho(0)K^\dagger_\beta]
          + [K_\alpha\rho(0),K^\dagger_\beta]\right),
\end{equation}
where 
\begin{equation}
S(t) = \frac{i}{2}\sum_{\alpha=1}
	[\chi_{\alpha 0}(t) K_\alpha - \chi_{0 \alpha}(t) K_\alpha^\dagger]. 
\end{equation}
For BB operations, a short-time expansion of 
Eq.~(\ref{eq:OSRexp}) is relevant. Choosing a hermitian operator 
basis $\{K_{\alpha }\}$, to first order in $\tau$ 
\begin{equation}
\rho(\tau ) \approx i [S(\tau ),\rho (0)],
\end{equation}
where $S(\tau )=\sum_{\alpha \geq 1}{\rm Im}(\chi^{(1)} _{\alpha 0}(\tau
))K_{\alpha }$, $\chi^{(1)}_{\alpha 0}(\tau) = 
\tau (d(\chi_{\alpha 0})/dt)_{t=0}$ and 
$K_0\equiv \Bid$ \cite{Lidar:CP01}.  Note that $S(\tau)$ 
behaves as a Hamiltonian.  Thus, using the abbreviation 
$\overline{\chi}_{\alpha}\equiv {\rm Im}(\chi^{(1)}_{\alpha 0}(\tau ))$, 
under the action of a group $\mathcal{G}=\{U_{k}\}_{k=1}^{N}$
of unitary BB controls $S(\tau)$ transforms as 
\begin{equation}
S(\tau) \rightarrow \sum_k U_k S(\tau)U_k^\dagger.
\end{equation}
Therefore the operator basis transforms as 
\begin{eqnarray}
\sum_\alpha \overline{\chi}_{\alpha}K_{\alpha } 
&\rightarrow& \frac{1}{N} \sum_\alpha \overline{\chi}_{\alpha}
              \sum_{k}U_{k}^{\dagger }K_{\alpha }U_{k} \nonumber \\ 
&&\;\; = \frac{1}{N}\sum_{\alpha \beta}\sum_k \overline{\chi}_{\alpha} 
        R_{\alpha \beta}^{(k)} K_{\beta }.
\end{eqnarray}
The last expression implies that $S$ and therefore 
$\chi$ transform according to the adjoint representation 
of $\mathcal{G}$, defined by $\sum_{\beta }R_{\alpha \beta
}^{(k)}K_{\beta }=U_{k}^{\dagger }K_{\alpha }U_{k}$.  For 
example $R\in SO(3)$ for $U\in SU(2)$, which leads to 
a geometric description of the result \cite{ByrdLidar:01}.  
Specifically,  we have under BB that $\sum_{\alpha \geq 1}
\overline{\chi}_{\alpha }K_{\alpha }\rightarrow \sum_{\beta\geq 1}
\tilde{\chi}_{\beta }K_{\beta }$, where 
\begin{equation}
\tilde{\chi}_{\beta }= \frac{1}{N}\sum_{k}\sum_{\alpha \geq 1}
\overline{\chi}_{\alpha}R_{\alpha \beta }^{(k)}.  \label{eq:BBtom}
\end{equation}
  Define 
$\hat{\chi}_\beta$  as the expansion coefficients of a `desired' 
Hamiltonian and the coefficients $\tilde{\chi}_{\beta }$, the 
BB-modified evolution.  E.g., for \emph{storage} the target evolution 
would be one for which all $\hat{\chi}_{\beta }$ vanish.  
For \emph{computation} we
would have a set of non-vanishing $\hat{\chi}_{\beta }$ 
describing the Hamiltonian we would wish to implement \cite{ByrdLidar:01}. 
{\em The key idea of empirical BB is to use the experimentally 
determined} $\overline{\chi}_{\alpha }$, 
{\em together with a specified set of} 
$\hat{\chi}_{\beta }$ {\em (corresponding to a 
desired evolution), to solve Eq.}~(\ref{eq:BBtom}) {\em for the rotation 
matrices} $R_{\alpha \beta }^{(k)}$ such that $\tilde{\chi}_{\beta } 
= \hat{\chi}_{\beta }$.  
These, in turn, determine a set of BB
operations \cite{ByrdLidar:01}. Thus, using the empirical BB method, 
\emph{one may
determine the required BB operations directly from experimental data}.  
In practice one would wish to minimize the difference between the 
target and BB-modified evolutions.  This difference can be described 
by any of the standard measures of distance including the Euclidean 
distance between the corresponding vector fields \cite{ByrdLidar:01}.  
Repeatedly performing the BB procedure determines the optimal BB process,
given the available controls and accounting for constraints, through a
control loop \cite{Judson:92}. In this manner only the experimentally
relevant errors are ever addressed, thus potentially reducing the size of
the set of BB operations.


\subsection{Bang-Bang Operations on Encoded Spaces}

Since the introduction of QECC \cite{Shor:95}, encoding 
techniques have become extremely important.  They have been 
used for DFSs \cite{Zanardi:97c,Duan:98,Lidar:PRL98} and 
universality considerations
\cite{Bacon:99a,Kempe:00,Bacon:Sydney,DiVincenzo:00a,Levy:01,Benjamin:01,WuLidar:01,LidarWu:01,Kempe:01,Kempe:01a,LidarWuBlais:02},
in some case combining DFS and QECC ideas \cite{Lidar:PRL99,Lidar:00b,KhodjastehLidar:02}.
Here we wish to take advantage of the benefits of encoding 
techniques while 
reducing noise in quantum systems using BB operations (see also 
\cite{Fortunato:01,Viola:01a,Boulant:02} for related results and ideas).  Indeed, 
it will be shown that BB operations on encoded operations can 
be very advantageous for the BB requirements as well.  
We believe that the methods for universal quantum computation and BB
controls using the $\{|01\rangle ,|10\rangle \}$ code (below) are of
immediate value to solid-state QC implementations. 
Let us first present a generally applicable result which 
gives sufficient conditions for the elimination of errors 
on encoded spaces using BB operations.  Let $\overline{\mathcal{G}}$
denote the generators of the group of logical operations (e.g., 
$\overline{\mathcal{G}}=\{\overline{X},\overline{Y},
\overline{Z}\}$, acting as {\em gates} on a single encoded qubit). 
In analogy to standard BB theory 
\cite{Viola:98,Duan:98e,Viola:99,Zanardi:98b,Vitali:99,Viola:99a,Viola:00a,Vitali:01,ByrdLidar:01,Cory:00,Agarwal:01,Uchiyama:02}
we
define ``symmetrization of a Hamiltonian $H$ with respect to 
$\overline{\mathcal{G}}$'' as:
$H\mapsto \sum_{U\in \overline{\mathcal{G}}}U^{\dagger }HU$.  
We then have the following result, which is a straightforward 
generalization of the BB condition for unencoded qubits 
\cite{Zanardi:99a,Viola:99}: 

\emph{Theorem 1}: Symmetrization with respect to $\overline{\mathcal{G}}$ 
suffices to completely decouple the dynamics of the encoded subspace.

\emph{Proof}: Symmetrization takes any system-bath Hamiltonian and 
projects it onto the centralizer of the group 
generated by $\overline{\mathcal{G}}$
(i.e., the set of elements that commutes with all elements of this group). 
By irreducibility of the 
representation of $\overline{\mathcal{G}}$, it follows, from
Shur's Lemma, that the BB-modified system-bath Hamiltonian is 
proportional to identity on the code space. I.e., the code space 
dynamics will be decoupled.  

This theorem shows that encoded BB operations 
may be combined with \emph{any encoding}. Of particular interest 
are DFSs and QECCs.  In addition, the sufficiency of the 
logical operations is important since they are 
assumed to be available in experiments.  
Later in this article we will discuss in detail a physically 
applicable case in which BB operations may be combined with a DFS.  
Here we briefly comment on how they may be combined with a QECC (see
also the experiment \cite{Boulant:02}).  

An obvious way in which BB operations may be used 
in conjunction with QECCs is the following: one may simply apply BB operations 
to each individual qubit.  This may well reduce the error 
rate and thus make an error correction code feasible when 
it would not be otherwise.  However, there are less obvious, 
but still beneficial techniques for combining these methods.  

QECCs can often be described by a stabilizer 
$\mathcal{S}=\{S_{i}\}$ \cite{Gottesman:97}, which is a group that has all
codewords as eigenstates with eigenvalue 1. The errors $\mathcal{E}
=\{E_{j}\} $ that a stabilizer code can detect are exactly the 
operators that \emph{anticommute} with
at least one element of $\mathcal{S}$ 
\cite{Gottesman:97}. To every stabilizer QECC there also corresponds a
set of logical operations (the normalizer), that is composed of
operators that commute with the stabilizer, and thus preserve the code
space. Every error $E_{j}$ also anticommutes with at least one of
element of the normalizer:
$\{\overline{g}_i,E_l\}=0$. This immediately implies that the logical
operations can be used as elements of an encoded BB control
scheme. The same is true for the elements of the
stabilizer. Thus, \emph{to suppress} $\mathcal{E}$, 
\emph{apply the generators of} 
$\mathcal{S}$ \emph{or of the normalizer as a set of BB operations}. Furthermore, since syndrome measurement for a stabilizer
code corresponds to measuring the elements of the stabilizer, 
BB operations can be applied during 
the measurement procedure. The final component of a QECC loop are
recovery operations, which typically correspond to applying the
inverse of the error operators. It is clear that therefore BB
operations cannot be applied during recovery, as they anticommute with
the recovery operations. Thus BB operations can be applied 
during the entire QECC procedure, with the exception of the 
recovery operations, without loss of the 
desired interaction.  (See \cite{ByrdLidar:01} for a geometric 
explanation of this.)  


For a demonstration of these considerations
consider the following simple, 
but important example of trying to protect against all single 
qubit errors.  The smallest QECC uses 5 physical qubits per logical 
qubit \cite{Laflamme:96}.  Instead, we could
start by encoding 1 logical qubit into 3: $|0\rangle _{L}=|000\rangle$,
$|1\rangle _{L}=|111\rangle$, in order to protect just against
independent bit flip errors $\mathcal{E}_{X}=\{X_{1},X_{2},X_{3}\}$ 
\cite{Nielsen:book} ($X_{i}$ represent the Pauli matrix $\sigma _{x}$ acting on the 
$i^{\mathrm{th}}$ qubit, etc.).  Given the conditions above, the 
logical operations come from the set 
$\{\overline{X} = X_1X_2X_3,\;\overline{Y} = -Y_1Y_2Y_3,\;
\overline{Z} = Z_1Z_2Z_3\}$.  The three qubit code 
leaves independent phase flip errors 
$\mathcal{E}_{Z}=\{Z_{1},Z_{2},Z_{3}\}$.  We can suppress these 
using the following BB operations on the
encoded qubits.  The stabilizer for the 3 qubit code for phase flips is
$\mathcal{S}_{X}=\{X_{1}X_{2},X_{2}X_{3},X_1X_3\}$, which clearly
anticommutes with $\mathcal{E}_{Z}$. Note that here $X_{i}X_{j}$ 
are gates, not Hamiltonians, and are therefore implemented using simultaneous
application of the single-body Hamiltonians $X_{i}$ and $X_{j}$.  Thus,
frequent application of the stabilizer elements as parity kick
operators will suppress the $\mathcal{E}_{Z}$ errors.  Since they 
are elements of the stabilizer, they will commute with the 
logical operations and thus, in principle, can be simultaneously applied.  
These stabilizer operations will leave no component of error 
in the $Y_i$ or $Z_i$ directions when implemented as BB.  
When one measures for 
$X_i$ errors, they will be projected onto the eigenbasis in which the 
measurement is performed.  This will not affect the $Y_i$ or 
$Z_i$ directions.  The advantage of these 
schemes, compared to the 5-qubit code \cite{Laflamme:96}, is in the
conservation of qubit resources.  
Of course, this comes at the expense of
additional gate operations which must be included in the QECC circuitry, 
but this may well be a worthwhile trade-off in 
situations where qubits are scarce.  We now turn to 
an explicit demonstration of combining DFS and BB to QC proposals 
based on exchange-type interactions.


\section{QC in Solid State Devices}

We now wish to discuss the application of the aforementioned techniques 
to quantum computing (QC) devices which use a form of the 
exchange interaction with particular emphasis on solid 
state proposals.  Essentially all promising solid-state QC proposals 
\cite{Fortschritte48} are based on either direct or effective 
exchange interactions between qubits, with a Hamiltonian of the form 
\begin{equation}
H_{\rm ex}=
\sum_{i<j}J_{ij}^{x}X_{i}X_{j}+J_{ij}^{y}Y_{i}Y_{j}+J_{ij}^{z}Z_{i}Z_{j}.
\label{eq:Hex}
\end{equation}
Representative examples are quantum dots 
\cite{Loss:98,Burkard:99,Hu:99,Imamoglu:99,Pazy:01,Levy:01a}, 
nuclear \cite{Kane:98} or electron \cite{Vrijen:00} spins of donor atoms in silicon, quantum Hall systems \cite{Mozyrsky:01}, and
electrons on helium \cite{Platzman:99}. These implementations combine
scalability with a clear route to controllability of qubit interactions via
tunable exchange couplings $J_{ij}^{\alpha}$. At the same 
time two major problems arise in
these proposals. Problem I: This, inherent problem, is shared by all other
QC proposals, and concerns the inevitable coupling to the
environment (lattice, impurities, and other degrees of freedom). This
coupling leads to decoherence, which introduces computational errors that
must either be prevented in the first place 
\cite{Zanardi:97c,Duan:98,Lidar:PRL98}, frequently corrected 
\cite{Shor:95,Steane:96a,Gottesman:97}, or suppressed 
\cite{Viola:98,Duan:98e,Viola:99,Zanardi:98b,Vitali:99,Viola:99a,Viola:00a,Vitali:01,ByrdLidar:01,Cory:00,Agarwal:01,Uchiyama:02}. Problem II: This, technological 
problem, is to some extent unique to solid-state QC architectures, and
concerns the fact that different constraints are involved in
implementing single-qubit versus two-qubit operations, for a 
variety of reasons detailed,
e.g., in \cite{WuLidar:01}. In fact the single-qubit operations often
involve significantly more demanding constraints. A large 
body of literature has been devoted to
overcoming the decoherence problem (for a review see \cite{Nielsen:book}),
some pertaining directly to quantum dots \cite{Burkard:99a}. A number of recent
papers have proposed solutions to the different constraints 
imposed by single and two
qubit operations \cite{Bacon:99a,Kempe:00,Bacon:Sydney,DiVincenzo:00a,Levy:01,Benjamin:01,WuLidar:01,LidarWu:01,Kempe:01,Kempe:01a,LidarWuBlais:02}.
Here, we propose a comprehensive and realistic solution
to both problems.


\subsection{Encoding}

We use a well-known code, first proposed in a quantum information
context in \cite{Palma:96}. Blocks of two 
qubits encode single logical qubits as follows: 
\begin{equation}
\left| 0_{L}\right\rangle _{i}\equiv \left| 0\rangle _{2i-1}\otimes
|1\right\rangle _{2i},\;\;\;\;\;\;\;\left| 1_{L}\right\rangle \equiv \left|
1\rangle _{2i-1}\otimes |0\right\rangle _{2i}.  \label{eq:encoding}
\end{equation}
Here $i=1,...,N/2$ indexes logical qubits, and $N$ is the total number of
physical qubits. It is simple to see how logic operations can be performed
on this code. Let us denote encoded logical operations by a bar; they\ act
on the encoded qubits in the same manner as the unencoded operations act on
physical qubits. E.g., $\overline{X}\left| 0_{L}\right\rangle =\left|
1_{L}\right\rangle $ and $\overline{X}\left| 1_{L}\right\rangle =\left|
0_{L}\right\rangle $. Then, the single-encoded-qubit logic operations,
defined by $\overline{X}_{i}=(X_{2i-1}X_{2i}+Y_{2i-1}Y_{2i})/2$ and $
\overline{Z}_{i}=(Z_{2i-1}-Z_{2i})/2$, viewed as controllable Hamiltonians,
can be used to generate all encoded-qubit $SU(2)$ transformations. Together
with the two-encoded-qubits operation $\overline{Z_{i}Z}_{i+1}
=Z_{2i}Z_{2i+1} $ that couples qubits in two neighboring blocks, and which
can be used to implement a controlled-phase transformation \cite
{Nielsen:book} between encoded qubits $i,i+1$, they form a \emph{universal
set of Hamiltonians} on the space of encoded qubits \cite{WuLidar:01}.
Universality means that by selectively turning the Hamiltonians $\{\overline{
X}_{i},\overline{Z}_{i},\overline{Z_{i}Z}_{i+1}\}$ on/off it is possible to
generate the Lie group $U(2^{N/2})$ of all possible transformations on the
encoded qubits.  Let us assume that the single-qubit spectrum is
non-degenerate, but not necessarily
controllable, i.e. the free Hamiltonian of the qubit system is 
$\sum_i\epsilon_i\sigma_i^z$, with $\epsilon_i\neq \epsilon_j$, 
but the $\epsilon_i$ are not separately tunable.  
As shown in \cite{LidarWu:01} it is then in fact sufficient to
actively control \emph{only} $\overline{X}_{i}$ in order to achieve
(encoded) universality, in the Heisenberg 
($J_{ij}^{x}=J_{ij}^{y}=J_{ij}^{z}$), XXZ 
($J_{ij}^{x}=J_{ij}^{y}\neq J_{ij}^{z}$), and XY 
($J_{ij}^{x}=J_{ij}^{y}$, $J_{ij}^{z}=0$) instances of the general exchange
Hamiltonian, Eq.~(\ref{eq:Hex}). The ``encoded recoupling'' method
introduced to this end in \cite{LidarWu:01} generalizes the standard NMR
selective recoupling method by applying pulses not to ``bare'' (physical)
qubits, but instead to encoded (logical) qubits. Encoded recoupling
eliminates the need for single-qubit control in exchange-based quantum
computer architectures and thus solves Problem II.

The second advantage of the above encoding is that it is a DFS with regard
to collective phase errors 
\cite{Palma:96,Duan:98,Lidar:PRL99,Kempe:00,Kwiat:00,Kielpinski:01}. 
Suppose the system is affected by a system-bath interaction Hamiltonian 
$H_{I}=S_{z}\otimes B_{z}$, where $S_{z}=\sum_{i}Z_{i}$ is the collective
dephasing operator. For logical qubit states $|\psi _{L}\rangle =a\left|
0_{L}\right\rangle +b\left| 1_{L}\right\rangle $, it is simple to check that 
$S_{z}|\psi _{L}\rangle =0$, so that $H_{I}$ does not affect the code. The
collective errors are expected to be particularly relevant for solid-state
systems at low temperatures and dephasing is one of the main problems in
this class of quantum computing devices. The DFS property of the above
encoding is therefore a partial solution to Problem I. However, collective
dephasing is by necessity an approximation. In realistic solid-state devices
there are other types of errors arising from a variety of sources. It is our
goal in this paper to show how the methods reviewed thus far can be extended
in a simple and realistic manner, to deal with these other sources of
decoherence. In particular, we now turn to the combination of these
techniques with the method of BB controls, except that, in the
spirit of encoded recoupling \cite{LidarWu:01}, we apply these controls on
the space of {\em encoded} qubits 
(see also \cite{Fortunato:01,Viola:01a}).  
The DFS encoding together with BB
operations on the encoded qubits will serve to counter decoherence, while
the method of encoded recoupling will allow for universal quantum
computation on the encoded qubits. The result of combining these 
three techniques
is the basis for our claim of a comprehensive solution to
problems of noise and design in solid-state quantum computing.


\subsection{Applying Bang Bang Operations on a Decoherence-Free Subspace}

As noted above, the logical qubits of Eq.~(\ref{eq:encoding}) are immune to
collective dephasing errors $Z_{2i-1}+Z_{2i}$. Let us focus on the first
encoded qubit ($i=1$), and consider which other errors can act on it. A
basis for all possible errors are the $2^{4}$ different tensor products of
all Pauli matrices (including the identity $I$) acting on two qubits.
Now, in general, four types of operations that
affect a DFS can be identified \cite{Lidar:PRL99}: (i) The set of 2 operations
to which the DFS is invariant -- ($I,Z_{1}+Z_{2}$); (ii) The set
of 3 operations that take states outside of the DFS to other states which are
also outside of the DFS. Both (i) and (ii) have no effect
on the DFS. (iii) The set of 3 logical operations -- [$\overline{X}
=(X_{1}X_{2}+Y_{1}Y_{2})/2,
\overline{Y}=(X_{1}Y_{2}-Y_{1}X_{2})/2,\overline{Z}=(Z_{1}-Z_{2})/2$]. 
When acting uncontrollably, these operations can cause 
\emph{logical} errors. (iv) The set of 8 operations which mix DFS states
with states out of the DFS -- see Eq.~(\ref{eq:errors}). These
operations are responsible for leakage from and into the DFS. Sets (iii) and
(iv) are those that damage the encoding. Both can cause decoherence by
entangling the encoded information with uncontrollable bath degrees of
freedom. Let us now apply this classification to our code. A basis for the
leakage errors (iv) is represented by the following set of operators: 
\begin{equation}
\{X_{1},X_{2},Y_{1},Y_{2},X_{1}Z_{2},Z_{1}X_{2},Y_{1}Z_{2},Z_{1}Y_{2}\}.
\label{eq:errors}
\end{equation}
This error set can clearly be seen to take the encoded states of 
Eq.~(\ref{eq:encoding}) out of the DFS (and vice versa) since it involves
single bit flips, or bit and phase flips on individual physical qubits.

We now come to a crucial observation first made in \cite{ByrdLidar:01a}. Let 
$U_{\overline{X}}(\phi)\equiv \exp (-i\phi \overline{X})$. 
Then, \emph{a single BB pulse of the form}
\begin{equation}
U_{\overline{X}}(\pi )=\exp (-i\pi
(X_{1}X_{2}+Y_{1}Y_{2})/2)=-Z_{1}Z_{2},
\end{equation}
\emph{can eliminate all type (iv) leakage errors}. That this is so
follows since $U_{\overline{X}}(\pi )$ anticommutes with all of the
errors in Eq.~(\ref{eq:errors}). As noted above [Eq.~(\ref{eq:PK})], 
this is the condition for
the parity kick version of BB controls. Thus, all type (iv) leakage errors
can be eliminated by a \emph{single} pair of BB pulses per 
cycle. This single pulse pair aspect is extremely important 
given the severe time constraints under which BB must operate.

In order to implement the BB operation $U_{\overline{X}}(\pi )$ it is
necessary to be able to switch on the Hamiltonian 
$J(X_{1}X_{2}+Y_{1}Y_{2})$ for a time $t=\pi /2J$.  
This (XY) Hamiltonian is directly available in a number of QC
proposals (quantum dots/atoms in cavities 
\cite{Imamoglu:99,Zheng:00}, quantum Hall systems 
\cite{Mozyrsky:01}). In dealing with systems that are governed by
the Heisenberg or XXZ Hamiltonians, the encoded selective recoupling
method can be used make these Hamiltonians simulate the XY 
type \cite{LidarWu:01}.
The Heisenberg case applies to the spin-coupled quantum dots 
and donor-spin proposals of \cite{Loss:98,Kane:98}. 
The XXZ case applies directly to the electrons on helium 
proposal \cite{Platzman:99},
and to the XY and Heisenberg proposals if symmetry
breaking mechanisms are taken into account 
\cite{WuLidar:01,LidarWu:01}. We note that
spin-orbit coupling induces anisotropic
terms that appear as corrections to Eq.~(\ref{eq:Hex}) \cite{Kavokin:01}. 
Methods for treating these have recently been suggested 
\cite{Bonesteel:01,Burkard:01,WuLidar:02}.  Thus the method using 
$U_{\overline{X}}(\pi)$ for eliminating leakage is applicable 
to a wide range of solid state QC proposals.

The elimination of all leakage errors by a single pair 
of BB operations per cycle is a rather
drastic alternative to the severe, factor of 5, qubit overhead incurred by
attempting to do the same using a concatenation of DFS and QECC encoding 
\cite{Lidar:PRL99}. The advantage is somewhat diminished if one is also
worried about the type (iii), logical, errors, which the DFS-QECC
concatenation method is capable of correcting at no extra cost \cite
{Lidar:PRL99}. Note first that $U_{X}(\pi /2)=-i\overline{X}$ 
anticommutes with both $\overline{Y}$ and $\overline{Z}$. 
Thus in fact \emph{all but one error} ($\overline{X}$ itself) 
\emph{can be
eliminated using just the single BB control Hamiltonian} $\overline{X}$. 
In order to
eliminate $\overline{X}$ as a logical error we must introduce other BB
controls, $\overline{Z}=(Z_{1}-Z_{2})/2$, and 
$\overline{Y}= i[\overline{Z},\overline{X}]$ 
which, by the theorem above, 
can then be used to eliminate other errors. To the
extent that these operations are available, this is a
reasonable proposition. However, since one of our goals was to avoid needing
to directly control single qubits, it is reassuring that the encoded
recoupling method \cite{LidarWu:01} can be used here again, in order to switch
on/off the Hamiltonian term $\overline{Z}$ by controlling $\overline{X}$
alone.

To summarize thus far, we have shown that the DFS encoding $|0\rangle
_{L}=|01\rangle $, $|1\rangle _{L}=|10\rangle $, which is immune to
collective dephasing errors, can be made robust against all leakage errors in
conjunction with the \emph{single} BB pulse $\exp (-i\pi \overline{X})$. 
To further eliminate all logical errors it is
necessary to introduce two more BB pulses, which can also be obtained 
from pulsing the XY Hamiltonian 
$\overline{X}=(X_{1}X_{2}+Y_{1}Y_{2})/2$.  This seems
like a modest set of requirements for the elimination of \emph{all}
decoherence errors on a single logical qubit, 
provided that the BB cycle time can indeed be made much
smaller than the bath timescale, as in Eq.~(\ref{eq:times}).  
We turn to an evaluation of this issue next, in the context 
of quantum dots.


\subsection{Estimation of Bath Cutoff Frequency in Quantum Dots}

Here we are primarily concerned with the spin-based GaAs quantum dots QC
proposals \cite{Loss:98,Burkard:99,Hu:99,Imamoglu:99}. The main spin 
relaxation and dephasing channels for
electron-spin qubits in GaAs have been recently thoroughly reviewed in 
\cite{Hu:01}. The dominant low temperature
mechanisms are related to spin-orbit coupling, which couples spins to
impurities and the lattice. Nevertheless, a lack of detailed understanding
of the various decoherence mechanisms persists. It is noteworthy that our
approach to error suppression does not rely on a detailed microscopic
understanding of these mechanisms. In the case of GaAs quantum dots, 
experimental
estimates for the spin dephasing time $T_{2}$ are $\sim 100ns$ 
\cite{Kikkawa:98}. We are not aware of direct measurements or
theoretical calculations of the bath cutoff frequency $1/\tau_{c}$ in these
systems. Nevertheless, we can provide positive evidence for the ability to
achieve the required BB pulse rates.

We consider the spin-bath and spin-boson models, which are rather general
models of low energy effective Hamiltonians, adaptable to a surprisingly wide
range of problems, including ours. The spin-boson model describes dephasing
due to coupling to delocalized modes (lattice vibrations), while the
spin-bath model captures the coupling to localized modes, such as nuclear
and paramagnetic spins, and defects \cite{Prokofev:00}. In
both models it can be shown that the characteristic decay time of coherence, 
$T_{2}=f(\tau _{c},T)$ ($\tau_c$ is the inverse of the bath 
spectral density high-frequency cutoff, $T$ is the temperature), 
and the function $f$ can be analytically determined in various cases 
\cite{Viola:98,Weiss:book,Palma:96,Lidar:CP01,Prokofev:00}. Note that
exponential coherence 
decay is rigorously valid only in the Markovian limit: e.g., in
the spin-boson model at $T=0$ with Ohmic damping, coherence decays
polynomially as $1/(1+(t/\tau _{c})^{2})$ \cite{Weiss:book}, 
in which case one can identify $
T_{2}=\tau _{c}$. In fact, since $\tau _{c}$ is the primary timescale
describing the bath, it is not unreasonable to quite generally identify $
T_{2}=c(T)\tau _{c}$, where $c$ is a function that depends only on $T$. This
is supported by a variety of instances of the spin-boson and spin-bath
models, differing by the specific form of the bath spectral density.
Furthermore, at low temperature $c(T)\approx 1$. Given $T_{2}\sim 100ns$ 
\cite{Kikkawa:98}, we thus conservatively estimate 
$\tau _{c} \sim 1-100ns$ for spin-coupled GaAs quantum dots. The gate
operation time in these systems is of the order of $50ps$ 
\cite{Hu:01}, and cannot be made much shorter because of induced
spin-orbit excitations \cite{Burkard:99}. Thus a range of $20-2000$ BB
parity-kick pulses seems attainable. The first order correction 
to the ideal limit of
infinitely fast and strong BB operations is $O((T_c/\tau_c)^2)$
\cite{Viola:98}, which, for parity kicks, in our case therefore 
translates to a correction of $O(10^{-2})$-$O(10^{-6})$.


\section{Conclusions}

To reduce noise and improve the reliability of quantum computing 
devices, new methods will have to be employed which take 
into account the constraints on current
experiments.  In 
particular, qubits are scarce resources today and will be 
in the near future.  In order to reduce qubit overhead in 
different error correction/avoidance encodings, 
we have presented two results for making the recently introduced 
BB operations more practical in present-day experiments.  The 
first is the theorem which 
gives necessary conditions for the removal of all errors on 
encoded spaces using logical operations alone.  The importance of 
this result lies in its generality; 
{\it Theorem 1} {\it gives sufficient conditions, in terms of 
logical operations, for the removal of all errors on 
an encoded subspaces via bang-bang controls using logical 
operations only}.  
Particularly important is the potential for {\it combining 
the error suppression methods of BB operations} 
\cite{Viola:98,Duan:98e,Viola:99,Zanardi:98b,Vitali:99,Viola:99a,Viola:00a,Vitali:01,ByrdLidar:01}, {\it with 
DFSs} \cite{Zanardi:97c,Duan:98,Lidar:PRL98} {\it and QECCs} 
\cite{Shor:95,Steane:96a,Gottesman:97},.  
The second result emphasizes the practical 
concerns of the experimentalist.  {\it Without relying on a particular 
model Hamiltonian we may, using quantum process 
tomography} \cite{Nielsen:book}, {\it determine an appropriate and 
efficient set of BB controls for 
physical and/or logical quantum computational states.}  

Application of our methods results in a rather comprehensive 
solution to problems of decoherence and gate
implementation in quantum computer proposals governed by exchange
Hamiltonians. Our solution combines ideas from the theory of
decoherence-free subspaces \cite{Zanardi:97c,Duan:98,Lidar:PRL98} 
and bang-bang (BB) controls 
\cite{Viola:98,Duan:98e,Viola:99,Zanardi:98b,Vitali:99,Viola:99a,Viola:00a,Vitali:01,ByrdLidar:01},
and the recently proposed method of encoded selective recoupling 
\cite{LidarWu:01}. By encoding logical qubits into pairs of physical qubits a
first level of protection against collective decoherence is obtained, which
can be further significantly enhanced using a \emph{single} 
type of BB operation,
that can eliminate all leakage errors from the DFS. Two more BB operations
are required to suppress \emph{all} other decoherence errors. We have
estimated that $10-1000$ parity-kick cycles can realistically be 
implemented in the
case of GaAs spin-coupled quantum dots within the bath correlation time.  
In conjunction with the elimination of the need for
difficult-to-implement single qubit operations enabled by the encoded
recoupling method \cite{LidarWu:01}, we believe that {\it our methods offer 
a realistic and comprehensive solution to some of the major 
difficulties associated with the design of quantum dot, and 
other exchange-based solid state quantum computers.}


\begin{acknowledgments}
We
thank Dr. L-A. Wu and Dr. K. Shiokawa for helpful discussions. This material is based on
research sponsored by the Defense Advanced Research Projects Agency under
the QuIST program and managed by the Air Force Research Laboratory (AFOSR),
under agreement F49620-01-1-0468 (to D.A.L.). The U.S. Government is authorized to
reproduce and distribute reprints for Governmental purposes notwithstanding
any copyright notation thereon. The views and conclusions contained herein
are those of the authors and should not be interpreted as necessarily
representing the official policies or endorsements, either expressed or
implied, of the Air Force Research Laboratory or the U.S. Government.
\end{acknowledgments}




\end{document}